\documentclass[%
reprint,
article,
amsmath,amssymb,
aps,
prl,
]{revtex4-2}

\usepackage{graphicx}
\usepackage{dcolumn}
\usepackage{bm}
\usepackage{mathrsfs}
\usepackage[mathscr]{euscript}
\usepackage{dsfont}
\usepackage[cal=boondox,scr=boondoxo]{mathalfa}
\usepackage{amsmath}
\usepackage{hyperref}

\begin{document}


\title{Anisotropic Special Relativity}

\author{Hassan Ganjitabar}%
\email{h.ganjitabar@ut.ac.ir}
\affiliation{%
	Department of Physics, University of Tehran, Amirabad, Tehran, Iran\\
	\\
}%

\date{\today}

\begin{abstract}
Anisotropic Special Relativity (ASR) is the relativistic theory of nature with a preferred direction in space-time. By relaxing the \textit{full-isotropy} constraint on space-time to the \textit{preference of one direction}, we obtain a perturbative modification of the Minkowski metric as $\mathscr{g}_{\mu\nu}\simeq\eta_{\mu\nu}+2\phi\epsilon_{\mu\nu}$ for a small perturbation parameter $\phi$. The symmetry group of ASR is obtained to have six generators satisfying the full Lorentz group algebra. However, the generators are deformed by the perturbation parameter $\phi$. So, ASR retains the same representations of Special Relativity (SR) but allows for Lorentz-invariant violation at the same time. Any invariant quantity of the theory is the inner product of two contravariant 4-vectors mediated by ${g}_{\mu\nu}$. The mass of a the particle is modified to $m^2=P^{\mu}\mathscr{g}_{\mu\nu}P^{\nu}$ which, in the first approximation level, has the extra term  $(2\phi\epsilon_{\mu\nu})P^{\mu}P^{\nu}$ compared to the mass of particle in SR. So, one application of ASR is, for example, to explain the neutrino flavour oscillation experiments in a natural way without violating the lepton number or adding sterile right-handed neutrinos. The mass of a particle is not the only quantity that is modified in ASR; any scalar quantity such as the Lagrangian of fields are also modified since the anisotropic metric $\mathscr{g}_{\mu\nu}$ is used to contract any pair of covariant-contravariant indices. As a more general consequence of ASR, any Quantum Field Theory (QFT) becomes anisotropic since the Lagrangian must contain the anisotropic metric. So, we provide a procedure to make anisotropic QFTs where the Lorentz-invariant Lagrangians are replaced with their ASR version.\\
\begin{description}
\item[PACS numbers]
11.10.-z, 03.30.+p, 11.30.Cp 
\end{description}
\end{abstract}

\pacs{Valid PACS appear here}
\maketitle


\section{Introduction}\label{sec1}
One of the widely accepted fundamental theories of physics is Special Relativity (SR) in which Quantum Field Theory (QFT) and Standard Model (SM) of the universe are grounded. In SR, quantities such as line element ${ds}^2=(dt^2-dx^2-dy^2-dz^2)$, are invariant under the Lorentz group transformations. However, despite the very strict restrictions on departure from Lorentz symmetry \cite{LIVTest1, LIVTest2, LIVTest3}, the Lorentz Invariance Violation (LIV) idea \cite{LIVIdea1,LIVIdea2} is still popular. One of the scenarios for LIV is Standard Model Extension (SME) \cite{SME} in which the Lorentz violation is allowed by adding non-relativistic terms to the Lagrangian. On the other side, there are theories wherein the LIV does not require the complete breakdown of relativistic symmetry. The examples of latter are perturbation approach of Coleman and Glashow \cite{Coleman} and Very Special Relativity (VSR) proposed by Cohen and Glashow \cite{VSR} both of which consider the anisotropy in space-time as an origin of LIV. So, the anisotropy of space-time is one of the possible scenarios for LIV, whose upper bound has already been investigated by analysing the cosmological data \cite{COSTest} as well as performing terrestrial experiments \cite{LASTest, SYNTest}.\\
There are two more theories allowing for LIV, Finslerian structure of the space-time \cite{FinslerBook} and Non-Commutative Quantum Field Theory \cite{NCQFT}, in context of which VSR can be realised \cite{VSR=Finsler, VSR=NCQFT}. In fact, a deformation of VSR symmetry can be seen in the case of a Finsler space-time in which the fundamental metric $g$ can be obtained from a function $F={(\eta_{\mu\nu}dx^{\mu}dx^{\nu})}^{(1-b)/2}{(n_{\alpha}dx^\alpha)}^{b}$ using $g_{\mu\nu}=\frac{1}{2}{\partial_\mu\partial_\nu}F^2$, where $\eta_{\mu\nu}$ is the Minkowski metric of SR, $n^{\alpha}=(1, 0, 0, 1)$ is a null vector and $b$ is a constant parameter. VSR suggests a subgroup of the Lorentz group, SIM(2), in addition to the space-time translations to be the maximal symmetry of nature wherein one of the three spatial directions is preferred. Also, on the track of building an anisotropic version of SR, Bogoslovsky suggested the Finslerian line element $ds={(\eta_{\mu\nu}dx^{\mu}dx^{\nu})}^{(1-b)/2}{(n_{\alpha}dx^\alpha)}^{b}$ which is not SR invariant but VSR invariant \cite{Bogoslovsky1, Bogoslovsky2}.\\
One of the main challenges for the aforementioned anisotropic theories is to allow for LIV and yet results into the same representations of the conventional Lorentz group. For instance, the symmetry group of VSR, SIM(2), has an abelian minimal group composed of two commuting generators $T_1=K_x+J_y$ and $T_2=K_y-J_x$. Therefore, it leads only to trivial $1\times1$ representations and does not provide a formalism consistent with the existence of fermion and bosons\cite{VSR=NCQFT}. 

In this paper, we propose a scenario wherein LIV is allowed through anisotropy without any need for deviation from the representations of the full conventional Lorentz group. We introduce a form of anisotropic metric that not only satisfies the anisotropy requirement but also leads to the same Lorentz algebra as SR does. On the one hand our theory allows for departure from Lorentz invariance while on the other hand it is consistent with the fermions and bosons fields. So, the challenge of having non-trivial representations in an anisotropic theory is well addressed in ASR.

An example of ASR consequences is modification of the mass dispersion relation that explains a natural mechanism for neutrinos to be massive without violating the lepton number or adding sterile right-handed neutrinos. So, this formalism can be considered in neutrino flavour oscillation experiments, a hot topic of high energy physics. Nevertheless, the impact of ASR goes beyond this example and could potentially affect any field theory since the anisotropic metric of ASR is needed to make ASR-invariant Lagrangians. We provide an instruction for making anisotropic QFTs; particularly, we outline the changes required to make an anisotropic version of Quantum Electrodynamics (QED).   

\section{Results and Discussion}\label{sec2}
We obtain anisotropic scalar quantities by easing the full-isotropy constraint on the Minkowski metric, but considering a preferred direction, $N^{\mu}$, in space-time instead. An anisotropic metric could be obtained either by varying the magnitude of the diagonal elements or introducing off-diagonal elements to $\eta_{\mu\nu}$. The minimal changes that can make the $z$-direction preferred without changing the determinant of metric is to have non-unity diagonal $t$- and $z$- elements. A modification of Minkowski metric as 
\begin{equation} \label{metric}
	\eta_{\mu\nu} \rightarrow \mathscr{g}_{\mu\nu}=\left(
	\begin{tabular}{cccc}
		$e^{2\phi}$ & $0$ & $0$ & $0$\\
		$0$ & $-1$ & $0$ & $0$\\
		$0$ & $0$ & $-1$ & $0$\\
		$0$ & $0$ & $0$ & $-e^{-2\phi}$\\
	\end{tabular}
	\right)
\end{equation}
induces such a desirable anisotropy. It is straightforward to check that the vector $N^{\mu}=(e^{-\phi}, 0, 0, e^{\phi})$ is a null-vector with respect to $\mathscr{g}$ ($\mathscr{g}_{\mu\nu}N^{\mu}N^{\nu}=0$); so, we consider the direction of $N^{\mu}$ as the preferred direction in space-time. As we have not varied any of the $x$- or $y$-elements of the metric, the isotropy is still retained in the $x-y$ plane. By Anisotropic Special Relativity (ASR) we mean a relativistic theory in which the inner product of a pair of covariant-contravariant indices is mediated by the anisotropic metric $\mathscr{g}$.

Applying metric (\ref{metric}) results in an anisotropic version of line element as
\begin{equation} \label{lineElement}
	{({ds}^2)}^{ASR}=(e^{2\phi}dt^2-dx^2-dy^2-e^{-2\phi}dz^2)
\end{equation}
\begin{equation} \label{lineElement1}
	{({ds}^2)}^{ASR}={({ds}^2)}^{SR}+2\phi(dt^2-dz^2)+O^2(\phi)
\end{equation}
which is not invariant under the conventional Lorentz symmetry but can be considered as a perturbation of the Lorentz invariant line element with the $\phi$ factor to be a constant parameter controlling the perturbation. Obviously, setting $\phi=0$ turns the situation back to the full Lorentz symmetry. The metric $\mathscr{g}$ itself can be also written as a perturbation of the Minkowski metric
\begin{equation}
	\mathscr{g}_{\mu\nu}=\eta_{\mu\nu}+2\phi\epsilon_{\mu\nu}+O^2(\phi)
\end{equation}
where
\begin{equation} \label{epsilon}
	\epsilon_{\mu\nu}=\left(
	\begin{tabular}{cccc}
		$1$ & $0$ & $0$ & $0$\\
		$0$ & $0$ & $0$ & $0$\\
		$0$ & $0$ & $0$ & $0$\\
		$0$ & $0$ & $0$ & $1$\\
	\end{tabular}
	\right),
\end{equation}
Moreover, it can be checked that the fundamental metric ${\mathscr{g}}$ introduced above is compatible with the full Lorentz group algebra. Starting from the line element invariance principal
\begin{equation} \label{ds2}
	{ds'}^2={ds}^2 \Rightarrow {\mathscr{g}'}_{\mu\nu}{dx'}^{\mu}{dx'}^{\nu}=\mathscr{g}_{\mu\nu}dx^{\mu}dx^{\nu}
\end{equation}
and considering $\Lambda^{\mu.}_{.\nu}$ as the most general transformation of the theory (excluding the space-time translations) that transforms the contravariant four-vectors such as ${dx}^{\mu}$ and the rank two tensors such as ${\mathscr{g}}_{\mu\nu}$ as:
\begin{equation} \label{transformations}
 {dx'}^{\mu}=\Lambda^{\mu.}_{.\nu}{dx}^{\nu}\,\,\,\,\,\,\,\,\&\,\,\,\,\,\,\,\,    {g'}^{..}_{\mu\nu}={\Lambda}^{.\rho}_{\mu.}{\Lambda}^{.\sigma}_{\nu.}{g}^{..}_{\rho\sigma}
\end{equation}
 one obtains
\begin{equation} \label{gequation}
  {\Lambda}^{\mu.}_{.\alpha}{\Lambda}^{.\rho}_{\mu.}{g}^{..}_{\rho\sigma}{\Lambda}^{.\sigma}_{\nu.}{\Lambda}^{\nu.}_{.\beta}={g}^{..}_{\alpha\beta} 
\end{equation}
By replacing $\Lambda^{\mu.}_{.\nu}=e^{({\Omega^{\mu.}_{.\nu}})}$ and focusing on infinitesimal transformations $\Lambda^{\mu.}_{.\nu}=\delta^{\mu.}_{.\nu}+\Omega^{\mu.}_{.\nu}$, Eq. (\ref{gequation}) results in
$\Omega^{..}_{\mu\nu}=-\Omega^{..}_{\nu\mu}$. Since $\Omega$ is antisymmetric, it has only six independent parameters as:
\begin{equation}\label{co-omega}
	\Omega^{..}_{\mu\nu}=\left(
	\begin{tabular}{cccc}
		$0$ & $\omega_{01}$ & $\omega_{02}$ & $\omega_{03}$\\
		$-\omega_{01}$ & $0$ & $\omega_{12}$ & $\omega_{13}$\\
		$-\omega_{02}$ & $-\omega_{12}$ & $0$ & $\omega_{23}$\\
		$-\omega_{03}$ & $-\omega_{13}$ & $-\omega_{23}$ & $0$\\
	\end{tabular}
	\right)
\end{equation}
The first covariant index of $\Omega^{..}_{\mu\nu}$ should now be converted to a contravariant one using $\mathscr{g}^{\mu\nu}$ (inverse of $\mathscr{g}_{\mu\nu}$) as $\Omega^{\mu.}_{.\nu}=\mathscr{g}^{\mu\alpha}\Omega^{..}_{\alpha\nu}$. In the matrix form, one need to calculate the inverse of metric $\mathscr{g}_{\mu\nu}$ given by Eq. (\ref{metric}) and apply it on the generator $\Omega^{..}_{\mu\nu}$ given by Eq. (\ref{co-omega}) to obtain:
\begin{equation}\label{omega}
	\Omega^{\mu.}_{.\nu}=\left(
	\begin{tabular}{cccc}
		$0$ & $\omega_{01}e^{-2\phi}$ & $\omega_{02}e^{-2\phi}$ & $\omega_{03}e^{-2\phi}$\\
		$\omega_{01}$ & $0$ & $-\omega_{12}$ & $-\omega_{13}$\\
		$\omega_{02}$ & $\omega_{12}$ & $0$ & $-\omega_{23}$\\
		$\omega_{03}e^{2\phi}$ & $\omega_{13}e^{2\phi}$ & $\omega_{23}e^{2\phi}$ & $0$\\
	\end{tabular}
	\right)
\end{equation}
$\Omega^{\mu.}_{.\nu}$ can be considered as the product of two antisymmetric tensors $\Omega^{\mu.}_{.\nu}=-\frac{i}{2}\omega^{..}_{\rho\sigma}{(J^{\rho\sigma}_{..})}^{\mu.}_{.\nu}$ with the expansion below:
\begin{equation}\label{Generators1}
	\begin{tabular}{cc}
		$\Omega^{\mu.}_{.\nu}=-i\omega^{..}_{01}{(J^{01}_{..})}^{\mu.}_{.\nu}-i\omega^{..}_{02}{(J^{02}_{..})}^{\mu.}_{.\nu}-i\omega^{..}_{03}{(J^{03}_{..})}^{\mu.}_{.\nu}$\\ $ \ \ \ \ \ \ \ \  -i\omega^{..}_{23}{(J^{23}_{..})}^{\mu.}_{.\nu}-i\omega^{..}_{31}{(J^{31}_{..})}^{\mu.}_{.\nu}-i\omega^{..}_{12}{(J^{12}_{..})}^{\mu.}_{.\nu}$
	\end{tabular}
\end{equation}
which can be rewritten using definitions:
\begin{equation}\label{Parameters}
	\left\{
	\begin{tabular}{cc}
		${(K^{i})}^{\mu.}_{.\nu}={(J^{0i}_{..})}^{\mu.}_{.\nu}$;     & $\xi^i=\omega^{0i}_{..}  $\\
		${(J^{i})}^{\mu.}_{.\nu}=-\frac{1}{2}\epsilon^{ijk}_{...}{(J^{..}_{jk})}^{\mu.}_{.\nu}$; & $\theta^i=-\frac{1}{2}\epsilon^{ijk}_{...}\omega^{..}_{jk}$
	\end{tabular}
	\right.
\end{equation}
as $\Omega^{\mu.}_{.\nu}=ie^{2\phi}(\vec{\boldmath{\xi}}.\vec{\boldmath{K}}-\vec{\boldmath{\theta}}.\vec{\boldmath{J}})$
where in 
\begin{equation}\label{InnerProduct}
	\left\{
	\begin{tabular}{c}
		$\vec{\boldmath{\xi}}.\vec{\boldmath{K}}=-g_{11}\xi^1K^1-g_{22}\xi^2K^2-g_{33}\xi^3K^3$\\
		$\vec{\boldmath{\theta}}.\vec{\boldmath{J}}=-g_{11}\theta^1J^1-g_{22}\theta^2J^2-g_{33}\theta^3J^3$
	\end{tabular}
	\right.
\end{equation}
By comparing Eq. (\ref{omega}) with Eq.(\ref{Parameters}), the six independent generators of the symmetry group, $K^i$ and $J^i$ can be found as:
\begin{equation} \label{Generators3}
	\begin{tabular}{cc}
	$\small{K^1}=\left(
	\begin{tabular}{cccc}
		$0$ & ${i}e^{-2\phi}$ & $0$ & $0$\\
		${i}$ & $0$ & $0$ & $0$\\
		$0$ & $0$ & $0$ & $0$\\
		$0$ & $0$ & $0$ & $0$\\
	\end{tabular}
	\right)$&
	$\small{J^1}=\left(
	\begin{tabular}{cccc}
		$0$ & $0$ & $0$ & $0$\\
		$0$ & $0$ & $0$ & $0$\\
		$0$ & $0$ & $0$ & ${i}e^{-2\phi}$\\
		$0$ & $0$ & ${-i}$ & $0$\\
	\end{tabular}
	\right)$\\
	$\small{K^2}=\left(
	\begin{tabular}{cccc}
		$0$ & $0$ & ${i}e^{-2\phi}$ & $0$\\
		$0$ & $0$ & $0$ & $0$\\
		${i}$ & $0$ & $0$ & $0$\\
		$0$ & $0$ & $0$ & $0$\\
	\end{tabular}
	\right)$&
	$\small{J^2}=\left(
	\begin{tabular}{cccc}
		$0$ & $0$ & $0$ & $0$\\
		$0$ & $0$ & $0$ & ${-i}e^{-2\phi}$\\
		$0$ & $0$ & $0$ & $0$\\
		$0$ & ${i}$ & $0$ & $0$\\
	\end{tabular}
	\right)$\\
		$\small{K^3}=\left(
		\begin{tabular}{cccc}
			$0$ & $0$ & $0$ & ${i}e^{-2\phi}$\\
			$0$ & $0$ & $0$ & $0$\\
			$0$ & $0$ & $0$ & $0$\\
			${i}e^{2\phi}$ & $0$ & $0$ & $0$\\
	\end{tabular}
	\right)$&
	$\small{J^3}=\left(
	\begin{tabular}{cccc}
		$0$ & $0$ & $0$ & $0$\\
		$0$ & $0$ & $i$ & $0$\\
		$0$ & $-i$ & $0$ & $0$\\
		$0$ & $0$ & $0$ & $0$\\
	\end{tabular}
	\right)$\\
	\end{tabular}
\end{equation}
It is straightforward to check that the above generators satisfy the full Lorentz group algebra:
\begin{equation}\label{algebra}
	\begin{tabular}{c}
		$[J^i, J^j]=+i\epsilon^{ijk}_{...}J_k$\\
		$[J^i, K^j]=+i\epsilon^{ijk}_{...}K_k$\\
		$[K^i, K^j]=-i\epsilon^{ijk}_{...}J_k$	
	\end{tabular}
\end{equation}
The fact that generators $K^i$ and $J^i$ satisfy the full-Lorentz group algebra warrants the representations of ASR to be the same as those of SR. On the other hand, the fact that ASR generators, given by Eq. (\ref{Generators3}), are deformations of Lorentz group generators means that ASR scalars do not need to be Lorentz invariant any more, see below. Unlike other existing theories, ASR explains LIV without leading to only trivial $1\times1$ representations \cite{NCQFT}; spinor and four-vector representations exist in ASR naturally. However, the equation of motion for such fields could vary as the fundamental metric $\mathscr{g}_{\mu\nu}$ deviates from the Minkowski metric.

As a consequence of ASR metric $\mathscr{g}$ given by (\ref{metric}), we can infer to introducing a Lorentz violating term into the fermionic Lagrangian to explain the neutrino's mass in a natural way without violating the lepton number or adding sterile right-handed neutrinos. This has been discussed for the first time in \cite{Neutrino, Bernardini}; however, here, a different Lorentz violating term, as a consequence of our metric $\mathscr{g}$ and the new deformed symmetry group, can  be added to the Lagrangian of neutrinos.  We consider the initially massless Majorana Lagrangian of neutrino and modify it just by applying metric $\mathscr{g}_{\mu\nu}$ to lower the index of one of the contravariant vectors e.i. $\mathcal{L}_\nu^{ASR}=i\bar{\nu}_M\gamma^{\alpha}\mathscr{g}_{\alpha\beta}\partial^\beta\nu_M$ which in the first order approximation leads to: 
\begin{equation}\label{NeutL}
	\mathcal{L}_\nu^{ASR}\simeq\mathcal{L}_\nu^{SR}+i\bar{\nu}_M(2\phi)(\gamma^0\partial^0+\gamma^3\partial^3)\nu_M
\end{equation}
where $\mathcal{L}_\nu^{SR}$ is the massless Majorana Lagrangian of neutrino in SR given by $i\bar{\nu}_M(\gamma^0\partial^0-\gamma^1\partial^1-\gamma^2\partial^2-\gamma^3\partial^3)\nu_M$ with $\nu_M$ being the four-component Majorana neutrino and $\gamma^{\alpha}$, $\alpha=0-3$, being the Dirac matrices. Eq. (\ref{NeutL}) gives the neutrino's mass dispersion relation as
\begin{equation}\label{NeutMass}
	{(m_\nu^2)}^{ASR}\simeq(p_\alpha p^\alpha)^{SR}+(2\phi)({(p^0)}^2+{(p^3)}^2)
\end{equation}
for which $(p_\alpha p^\alpha)^{SR}={(m_\nu^2)}^{SR}$ is assumed to be zero; even in such a case, it is still possible for $(m_\nu^2)^{ASR}$ to be non-zero due to the appearance of the perturbative term $(2\phi)({(p^0)}^2+{(p^3)}^2)$. However, for the neutrinos moving purely in the $x-y$ plane ($p^0=p^3=0$), the ASR mass yet remains zero. Therefore, the neutrino's Majorana mass in anisotropic space-time will depend on the direction of its movement which in turn means that the neutrino flavour oscillation allowed by anisotropic theories like ASR and VSR \cite{ElectroweakVSR} could be direction-dependent as well.

It can also be noticed that the field theory (FT) obtained in ASR remains \textit{local} in contrast to the non-local FT emerging from the correction term introduced by Cohen and Glashow in VSR \cite{Neutrino}. The term they suggested to correct the equations of motion with is $\frac{m_\nu^2\eta_{\alpha\beta}\gamma^\alpha n^\beta}{2n.\partial}\nu_M$ where the appearance of $\partial$ operator in the denominator makes the theory non-local. However, ASR-FT remains local as only the lower orders of the derivative operator appear in the Lagrangian.

In general, the mass of any field in ASR can be modified just similar to the line element in Eq. (\ref{lineElement1}) i.e.
\begin{equation}\label{Mass}
	{(m^2)}^{ASR}={(m^2)}^{SR}+(2\phi)({(p^0)}^2+{(p^3)}^2)+O^2(\phi)
\end{equation}
where $m^{SR}$ is the mass of the field in SR given by ${(m^{SR})}^2=({(p^0)}^2-{(p^1)}^2-{(p^2)}^2-{(p^3)}^2)$ with $p^i$, $i=0-3$, being the elements of the contravariant four-momentum vector in SR.

In order to make an anisotropic quantum field theory, our procedure is to promote any Lagrangian by replacing the isotropic inner product of any pair of covariant-contravariant indices with its anisotropic version. Similar to the Majorana Lagrangian of left-handed neutrinos, the Dirac Lagrangian for the fermions can be modified to
\begin{equation}\label{Lagrangian}
	\mathcal{L}_D^{ASR}\simeq\mathcal{L}_D^{SR}+i(2\phi)\bar{\psi}(\gamma^0\partial^0+\gamma^3\partial^3)\psi
\end{equation}
a consequence of which is the extension of the dispersion relation as (\ref{Mass}). Similarly, in the electrodynamics Lagrangian $\mathcal{L}_{EM}^{SR}=-\frac{1}{4}F^{..}_{\alpha\beta}F^{\alpha\beta}_{..}$ we just need to replace the Minkowski metric with $\mathscr{g}$ in lowering the indices.
\begin{equation}\label{EM_Lagrangian}
	\mathcal{L}_{EM}^{ASR}=-\frac{1}{4}(\mathscr{g}_{\alpha\mu}\mathscr{g}_{\beta\nu}F^{\mu\nu})F^{\alpha\beta}
\end{equation}
resulting in equations of motion as
\begin{equation}\label{E_EoM}
	(\partial_\mu F^{\mu\nu})^{ASR}\simeq(\partial_\mu F^{\mu\nu})^{SR}+(2\phi)\epsilon_{\mu\alpha}\partial^\alpha F^{\mu\nu}=0
\end{equation}
from which the anisotropically modified Maxwell equations can be derived.
The shape of gauge symmetry in ASR remains the same as in SR
\begin{equation}\label{Gauge}
	\left\{
	\begin{tabular}{c}
		$\psi\,\,\rightarrow\,\,e^{iq\vartheta}\psi$\\
		$\partial_\mu\,\,\rightarrow\,\,D_\mu=\partial_\mu-iqA_\mu$\\
		$A_\mu\,\,\rightarrow A_\mu+\partial_\mu\vartheta$
	\end{tabular}
	\right.
\end{equation}
with the difference that, here, all the covariant indices are made from the contravariant ones by applying the metric $\mathscr{g}$. For instance, the ASR version of $A_\mu$ is
\begin{equation}
	A_\mu^{ASR}=\left(
	\begin{tabular}{c}
		$e^{2\phi}A^0$\\
		$-A^1$\\
		$-A^2$\\
		$-e^{-2\phi}A^3$
	\end{tabular}
	\right)
\end{equation}
\section{Conclusion}\label{sec3}
In this paper, we proposed an anisotropic relativistic theory, ASR, by replacing the Minkowski metric with an anisotropic metric in which the space-time isotropy constraint has been replaced with the preference of a direction determined by the null vector $N^\mu$. On the one hand, a perturbative deviation of Lorentz invariance is allowed while on the other hand, the symmetry group's deformed generators satisfy the same algebra as that of the full-Lorentz group, which guarantees the irreducible representations of ASR to be just the same as those of SR. Despite a direction-dependent modification to the dispersion relation, the QFT built up in ASR remains local.\\
As one of the interesting applications of ASR, we can infer to redesigning of neutrino oscillation experiments. ASR suggests the neutrino mass, hence the neutrino oscillation, to be direction dependent; so, designing a neutrino oscillation experiment in which the direction of propagation of neutrinos can sweep different spatial directions could be of special interest. In addition, as the transformations of ASR are direction dependent, it is expected that most of the relativistic phenomena like length contraction, time dilation, Doppler effect and aberration of light, etc. to be direction dependent too. From a broader point of view, any QFT becomes anisotropic since the Lorentz invariant Lagrangians are modified to their ASR version. With the examples of Majorana, Dirac, and electromagnetic fields, we demonstrated how replacing the Minkowski metric with the anisotropic metric of ASR modifies the Lagrangian and hence the form of the equations of motion.\\
A possible origin for the space-time anisotropy could be the anisotropic mass/energy distribution in the universe, which can be reflected in cosmic microwave background. This means that different locations in the universe could be expected to experience different level or direction of anisotropy. In such a case, a universal version of ASR can be achieved by promoting the perturbative parameter $\phi$ from a constant to a field.\\

\bibliography{ASRCitation}
\bibliographystyle{nature}

\end{document}